# Even Small Companies Can Save Lives by Reducing Emissions

**Daniel Baldassare[1], Abby Lute[1], Hikari Murayama[2], Cora Kingdon[2], Christopher Schwalm[1]**

- [1]Woodwell Climate Research Center, Falmouth, MA 02540, USA
- [2]Energy and Resources Group, University of California, Berkeley, CA 92720, USA

Correspondence: Daniel Baldassare (dbaldassare@woodwellclimate.org)

## Significance Statement

Global warming efforts are often framed around broad thresholds, for example the 1.5°C target. While these targets are important, this focus can make it seem like small emissions cuts are meaningless if we fail to keep warming below 1.5°C. Here, we model avoided warming from the sustainability pledges of 3,084 companies to show the importance of small, independent emissions reductions. By using historical relationships between temperature and mortality, we estimate the life-saving potential of corporate pollution reductions at companies ranging from small startups to major oil producers. We find that over 92% of companies can save at least one life by implementing their sustainability pledges, with over 4.4 million lives saved across just these 3,084 companies.

**Abstract**

Global warming is often framed in broad planetary numbers such as the 1.5°C and 2°C warming thresholds, creating the false impression that individual corporations' efforts to reduce emissions are meaningless in the absence of collective action. This perspective causes companies to reduce ambition towards voluntarily cutting emissions, as they believe their pollution has negligible impacts on its own. Reframing the issue to focus on the life-saving potential of independent corporate actions empowers companies to act and holds them accountable for inaction. Here, we show the results from an innovative climate-health modeling technique which calculates the avoided deaths from sustainability efforts for 3,084 companies spanning a range of sizes and sectors. From the reported emissions and planned emissions reductions, we create scenarios for 2020-2049 with and without companies' pledged emissions cuts and calculate the resulting warming from 2020-2100 using a climate emulator. We then use temperatures from these scenarios to calculate the deaths resulting from warming by using mortality damage functions. We find that more than 92% of these companies stand to save at least one life by following through with emissions reduction plans. Additionally, if all 3,084 companies follow through with their emissions reduction plans, over 4.4 million temperature-related deaths can be avoided.

## Introduction

Global warming is typically framed in various thresholds such as the 1.5°C warming threshold, with failure and success defined as meeting or exceeding these warming thresholds. While this framing is useful for international coordination, it inadvertently creates the false impression that smaller emissions reductions are meaningless, particularly as warming thresholds are exceeded (Lamb et al., 2020; Painter et al., 2023; Kopp et al., 2024). Additionally, this narrative abdicates responsibility from individual polluters, reducing agency and ambition (Buhr et al., 2023; Privato et al., 2024). Shifting focus to the benefits of incremental efforts unlocks massive emissions reduction potential from smaller businesses and applies pressure to voluntarily reduce pollution (Boon-Falleur et al., 2022; Constantino et al., 2022; Herweg and Schmidt, 2022; Dewatripont and Tirole, 2023; Buettner et al., 2025). The need for alternate, non-mandated mitigation is particularly acute, considering a decreased appetite for regulation and following the recent weakening of national decarbonization efforts (Immink et al., 2021; Harris, 2022; Fransen et al., 2023). However, a current lack of information regarding benefits from corporate decarbonization hampers efforts. To this end, we use reported emissions and planned emissions reductions from a wide range of companies to compute the avoided warming and resulting health impacts from corporate decarbonization efforts, demonstrating that even smaller companies can save lives by reducing emissions.

Although implementation of corporate emissions pledges has been inconsistent (Dahlmann et al., 2019; CDP, 2025), voluntary corporate emissions cuts present the opportunity to drive significant reductions (Krabbe et al., 2015). As shown by the success of cap-and-trade, companies have proprietary information on individual decarbonization options that regulators do not possess, allowing companies to be uniquely well-positioned to drive emissions reductions, given sufficient internal and external pressures (Schmalensee and Stavins, 2016; Robiou du Pont et al., 2024). With enough flexibility, regulatory pressure has been found to increase rates of green innovation, lowering the cost of abatement (Lim and Prakash, 2023). In addition to pressure from regulators and other outside sources including investors and consumers, internal forces are major drivers of corporate sustainability and decarbonization programs (Gouda et al., 2020; Bello-Pintado et al., 2023). Internal forces can include senior leadership, sustainability teams, managers, and individual contributors, each of which possess unique abilities and knowledge to drive corporate sustainability efforts (Stoughton and Ludema, 2012). Internal engagement with sustainability is partially mediated by employees' understanding of sustainability impacts (Manninen and Huiskonen, 2022), creating a need for research on the benefits of small, uncoordinated decarbonization efforts in order to unlock the massive potential of voluntary corporate actions (Damert and Baumgartner, 2017).

The focus on keeping warming below 1.5°C, due to both the coarseness of the objective and the extremely low likelihood of its achievement, creates invisible barriers for voluntary decarbonization (Davidson, 2024; Kopp et al., 2024). This threshold has been described as a critical level of warming, below which the effects are manageable, and above which catastrophic and irreversible changes will occur in the climate system (Masson-Delmotte et al., 2019; Calvin et al., 2023; Ripple et al., 2024), downplaying the importance of incremental emissions cuts. This is evidenced by high-profile research emphasizing the catastrophic impact of "tipping points" at the 1.5°C warming level and suggesting, for example, that the climate-related damages in a 2°C and 3°C world are only "somewhat different" (Armstrong McKay et al. 2022). While research has continued to explore the nuances of damages in a below 1.5°C world (e.g. Estrada and Bozen, 2021) and the uncertainty of timing associated with "tipping points" (Ritchie et al., 2021), a layperson is unlikely to have been exposed to this discussion. Emphasizing the importance of keeping warming below 1.5°C may have increased ambition initially, but as a 1.5°C world becomes inevitable (Matthews et al., 2020; Matthews and Wynes, 2022), this focus presents the problem as intractable, hampering mitigation efforts (Bostrom et al., 2018; Xiang et al., 2019; Bradley et al., 2020; Roeflsma et al., 2020; Geiges et al., 2020; Calvin et al., 2023; Cherry et al., 2024; Kopp et al., 2025). As a field tackling a critical societal issue where framing can impact ambition, informing and motivating the public is an unavoidable reality (Schmidt, 2015) necessitating thoughtful communication.

In addition to the issues around already and soon-to-be exceeded warming levels, the large differences between emissions scenarios further limits the perceived agency of polluters (Baldassare and Reichler, 2024). The limited number of emissions scenarios (Riahi et al., 2016) or warming levels typically used in analyses of climate impacts, for example in recent IPCC reports (Masson-Delmotte et al., 2019; IPCC, 2021; Calvin et al., 2023), reinforces the idea that individual corporations' emissions are unimportant, as even the largest polluters are incapable of reducing warming enough to bridge the gap between these scenarios (Callahan and Rankin, 2025). The scenarios are so disparate that only concerted international coordination appears to be capable of meaningfully reducing warming, giving companies the cover to wait for regulation that may never come. This results in a challenging environment for corporate sustainability, where the costs can be easily calculated but the benefits of decarbonization efforts are unclear.

Currently, multiple approaches exist tying individual emissions to either current or future climate damages. One prominent metric, commonly used in decision-making, is the Social Cost of Carbon (SCC), which is calculated through econometric damage functions (Ricke et al., 2018) and represents the cost to society, in dollars, that occurs from each additional tonne of emissions.

This currency-denoted metric has increased substantially in value since its inception as more climate damages are incorporated into SCC modeling efforts (Rennert et al., 2022; Tol, 2023). The SCC is used to account for the societal harm of greenhouse gas emissions in regulatory processes in a few countries and several U.S. states, including Canada, Germany, California, Colorado, Minnesota, New York, and Washington. While useful for regulation and analysis, in the absence of regulator-mandated consideration, responsibility for discounted economic damages over the next few centuries is unlikely to be compelling for companies facing more immediate pressures from competitors and customers (Peters et al., 2022; Peters and Salas, 2022; Lenton et al., 2023). The present situation, with SCC as the only viable quantification of uncoordinated decarbonization benefits, hampers corporate ambition, necessitating improved methods to quantify the importance of independent corporate sustainability efforts.

To this end, extreme event attribution research has emerged, which aims to estimate the impact of greenhouse gas emissions on specific extreme weather events (Swain et al., 2020). This field has connected a wide range of extreme events, from wildfires (Law et al., 2025) to hurricanes (Reed and Wehner 2023) and heatwaves (Ignjačević et al., 2024) to anthropogenic emissions. Recent studies have extended this research to estimate the responsibility of individual major polluters for heat-related economic losses (Callahan and Mankin, 2025). This attribution research is further buttressed by quantitative analyses of the health impacts of particulate emissions associated with greenhouse gas emissions (Dinh et al., 2024; Stewart et al., 2025; Zhao et al., 2025). However, currently these methods are limited to the largest polluters, preventing the quantification of the impacts from smaller polluters' emissions or decarbonization efforts.

An additional method for visualizing negative consequences of emissions uses observed relationships between temperature changes and excess deaths (Gasparrini et al., 2015; Gasparrini et al., 2017; Lee et al., 2020; Bressler et al., 2021; Cromar et al., 2022). This method allows for an estimation of changes in heat and cold-related deaths in a warming climate, providing a crucial link between emissions and impacts. By presenting global warming damages in deaths rather than dollars, these estimates are likely to be more compelling to polluters than Social Cost of Carbon estimates (Peters and Salas, 2022). At present, this method is limited to historical relationships between warming and heat and cold-related deaths and therefore does not account for the impacts of intensifying storms, chronic damages, or secondary effects such as increased conflict. Additionally, there is emerging research on increased mortality from growing wildfire smoke exposure due to climate change (Qiu et al., 2025), which is not included in these temperature-related mortality estimates (Cromar et al., 2022). As such, these temperature-related mortality estimates likely represent a lower bound of global warming induced mortality, as they focus on only one direct threat among many direct and indirect threats.

Here, we project the avoided deaths from individual companies' emissions reduction targets by using recent advances in estimating the temperature-related deaths resulting from greenhouse gas emissions. This analysis expands the quantifications of impacts from attribution science to smaller polluters and clarifies the benefits of decarbonization efforts at a wide range of companies. We obtained reported emissions and stated emissions reduction targets from Tracenable, a corporate sustainability data provider, for a wide range of companies (Tracenable, 2025). These emissions were used to generate scenarios in which planned emissions reductions are implemented, delayed, or where no reductions occur. These scenarios were run through a climate emulator, resulting in global warming estimates, which were then used with global mortality damage functions from Wells et al. (2025), adapted from Bressler et al. (2021) and Gasparrini et al. (2017) to estimate the lives saved from corporate emissions cuts. As we show, the vast majority of companies are able to save at least one life by following through with emissions reductions, with over 4.4 million heat-related deaths across all companies. These methods are also adapted to custom scenarios, creating a framework for visualizing the life-saving potential of decarbonization which we hope can drive voluntary corporate emissions cuts.

**Methods**

In order to estimate the mortality reduction from corporate emissions pledges, we combine reported emissions with stated emissions reduction targets to create future emissions scenarios, which are then used in a climate emulator to produce future warming estimates that are used in mortality damage functions. Reported annual Scope 1 and 2 $CO_{2e}$ emissions for 2020-2023 were acquired for 7,635 companies from Tracenable, a corporate sustainability data provider (Tracenable, 2025). Scope 1 emissions are those that are directly produced by the company, Scope 2 emissions are produced by energy purchased by the company, and Scope 3 cover a wide range of direct and indirect emissions associated with operations. Scope 3 emissions were excluded due to a relative scarcity of reported Scope 3 emissions and publicly stated emissions reductions plans for Scope 3 emissions. Tracenable was selected as the data provider as they provide both reported emissions and reduction targets using a unique AI with human oversight approach to produce sustainability data which can be traced to the original source documents. The 7,635 companies represent all of the companies for which Scope 1 or 2 emissions data was available on Tracenable on September 16, 2025. The companies range from small polluters with less than 1,000 tonnes of annual $CO_{2e}$ emissions to oil producers with over 100,000,000 tonnes of annual $CO_{2e}$ emissions, greater than the 2023 emissions of over 150 countries (Crippa et al., 2024). Stated Scope 1 and 2 emissions reduction pledges were acquired from Tracenable for 3,903 companies, representing all companies available on Tracenable on September 16, 2025. Included emissions reductions consist of a baseline year, a target year, and a reduction percentage, resulting in the removal of companies with only intensity-based emissions reduction targets. As only companies with emissions reduction plans in the Tracenable database are included here, the analyzed companies are an imperfect representation of global companies both in geographic distribution and sector (see Supplement Figure S1). The analyzed companies are disproportionally based in Asia and Europe, in line with recent Science Based Targets Initiative reporting of corporate emissions target setting (SBTi, 2025).

Reported emissions data was subjected to quality control, removing companies with non-numeric emissions, or where emissions are clearly unrealistically large. Scope 1 and 2 emissions were combined to form reported emissions. Emissions reductions intentions were subjected to similar quality control, removing non-numeric pledges, pledges in units other than percent reduction, or pledges greater than 100%. For emissions reductions, where multiple records exist for a single company, often due to the presence of separate Scope 1 and 2 emissions reduction pledges, the lowest of the values were used and the others were discarded. Companies with missing emissions for any years between 2020 and 2023 were filled with the company's average emissions across existing data.

For all of the companies with pledges, a baseline No Emissions Reductions scenario for 2020-2049 was created by projecting continuous 2023 $CO_{2e}$ emissions out to 2049. Two additional scenarios were created for each company, a Stated Emissions Reductions scenario where a company fully implements its emissions reduction pledges, and a Delayed Emissions

Reductions scenario where pledges are fully implemented 5 years after the stated date. While it is impossible at present to estimate how actual reductions will compare to future targets, the 5-year Delayed Emissions Reductions scenario presents an alternate future based on the present reality that many companies have already failed to meet their targets (Aldy et al., 2023). The period from 2020-2049 is selected as a compromise between the higher confidence of a shorter time period and the longer time horizon required for decarbonization of carbon-intensive processes. Target dates in the dataset range from the early 2020s to 2060s, indicating the wide range of timelines associated with decarbonization efforts (see Supplementary Figure S2). Pledged reductions were converted into annual emissions reduction percentages by dividing the reduction percentage by the number of years between the target year (or the target year plus 5 for the Delayed Scenario) and the baseline year. For each year, the annual emissions reduction percentages were multiplied by the year minus the baseline year for 2024-2049, which was then multiplied by the baseline scenario to create the emissions data in the Stated and Delayed Emissions Reductions scenarios. Emissions reductions were capped at 100%, preventing emissions from becoming negative.

Additional scenarios were created for each company using defined emissions reduction rates of 1%, 2%, 5%, and 10% per year beginning in 2024 and continuing through 2049. This resulted in four additional scenarios from 2020-2049 for all companies for which reported $CO_{2e}$ emissions were available. These prescribed emissions reduction scenarios are used to determine the likelihood that a given annual emissions reduction rate for a polluter of a certain size would result in mortality reductions, allowing for an analysis of mortality reduction potential across these companies. In comparison to the previous scenarios, these scenarios do not consider decarbonization ambition but instead solely focus on impacts following various decarbonization strategies. Similar to the other scenarios, emissions reductions were capped at 100%. Although it would have been possible to include companies without emissions reduction pledges in this analysis, for comparison with the previous analysis only companies with emissions reduction pledges were included. This resulted in a total of 3,084 companies analyzed in each scenario.

To analyze the warming implications from various emissions reductions, the Finite Amplitude Impulse Response (FaIR) climate emulator version 2.1.3 (Leach et al., 2021) was used. FaIR is an ultra-lightweight climate model, enabling efficient modeling of global temperatures given prescribed greenhouse gas forcings. Greenhouse gas forcings were obtained from the Network for Greening the Financial System (NGFS); for efficiency only the NGFS Current Policies scenario (NGFS, 2024) was used, which projects emissions following only emissions reduction policies which have already been announced. One run was conducted with the Current Policies as inputs, which is the baseline scenario against which the company-specific simulations were compared. To compute the warming for each company in each scenario, emissions for the given company in the given scenario were subtracted from the Current Policies Scenario's $CO_{2e}$ from fossil fuel combustion and industrial processes. In this way, the scenarios represent a simulation where the company reduced its emissions by a certain amount, and the difference between the Current Policies scenario and the company-specific scenario represents the companies' contribution to warming. The decrease in projected warming resulting from emissions reductions for a given company and scenario was calculated by subtracting the warming in the No Reduction scenario, where emissions from 2024-2049 are equal to 2023 emissions, from the given scenario with reduced emissions. The result of this analysis is a table of temperature anomalies relative to 1750 for each year from 2020-2100 for each company in each of the seven scenarios.

To calculate the potential increase in deaths from resulting warming, mortality damage functions were applied to the FaIR temperature outputs. The global mortality damage functions, obtained from Wells et al. (2025), which are adapted from the country-specific functions from Bressler et al. (2021), estimate the net change in temperature-related mortality, including both the deaths resulting from increased high temperatures and the avoided deaths from warmer low temperatures. The mortality-temperature damage functions build off Gasparrini et al. (2015) and Gasparrini et al. (2017) which estimate changes to heat and cold-related mortality using historical daily temperature and all-cause mortality data, and are similar in value to the damage functions from Cromar et al. (2022). These damage functions do not account for deaths from intensified storms, droughts, or secondary impacts from these events. Additionally, sea-level-rise and other chronic hazards are not included in the mortality estimates. Also omitted are the co-benefits of reduced pollution, for which a wide range of estimates exists in the literature (Ngan Thi Thu et al., 2024). While this approach leaves significant room for improvement as it is solely focused on temperature-related mortality, we chose to implement these mortality damage functions as they provide a reasonable lower-bound estimate of the mortality impacts of emissions.

These mortality damage functions relate temperature to all-cause mortality ratios, necessitating all-cause mortality projections. For this, we chose global annual death projections from the United Nations World Population Prospects 2024 (United Nations, 2024) which contain probabilistic forecasts of total deaths from 1950-2100. Because we are interested in companies' impacts from reported and projected emissions, we utilized data from 2020 to 2100, the final year available in the all-cause mortality dataset. While emissions are only projected from 2020-2049 due to increasing uncertainties around emissions beyond 2050, mortality was measured from 2020-2100 to account for the long-lasting effects of emissions. This resulted in a significant undercount of mortality impacts from corporate emissions as temperature-related deaths after 2100 were not included in the analysis.

Mortality for each year in each scenario was calculated by multiplying the annual temperature by the mortality damage function and the total deaths for that year. As we are interested in the mortality caused by a given company's emissions, this value was then subtracted from the NGFS Current Policies baseline run, resulting in the annual mortality caused by the company's emissions. Reductions in mortality were calculated by subtracting the mortality from one of the reduced emission scenarios from the No Reduction scenario. While the probabilistic FaIR simulations can allow for an analysis of warming uncertainty, we focused only on the mean estimate for each company due to a lack of data regarding the uncertainty of baseline corporate emissions for 2024-2049 and efficiency considerations given the large number of simulations.

## Results

We begin by analyzing the emissions cuts in 3,084 companies' stated emission reduction targets. Each company provides a baseline year, target year, and percent emissions reduction, allowing for the calculation of an annual emissions reduction rate assuming emissions reductions occur at a steady pace. For each of these 3,084 companies, we create a No Emissions Reduction baseline emissions scenario assuming constant 2023 emissions through 2049. We also create a Stated Emissions Reductions scenario, which assumes that the annual percent reductions associated with companies' emissions targets continue through 2049. The third scenario we define is a Delayed Emissions Reductions scenario, which is identical to the previous scenario, but with the target year delayed by 5 years. For the Stated and Delayed Emissions Reductions scenarios we calculate the cumulative emissions reduction percentage relative to the no reduction scenario (Figure 1). Percent reductions in 2020-2049 emissions are largest for smaller polluters, indicating that larger polluters are less likely to plan for aggressive emissions reductions. Regardless of these differences, implementation of pledged emissions reductions would result in a greater than 50% reduction of 2020-2049 emissions for most companies, and a 5-year delayed target still results in reductions greater than 35% for most companies as shown by the 95th percentile confidence interval in Figure 1. Smaller polluters have the largest range in emissions reductions, with emissions reductions from 25% to nearly 100%, suggesting a wide range of capabilities and interest amongst smaller polluters.

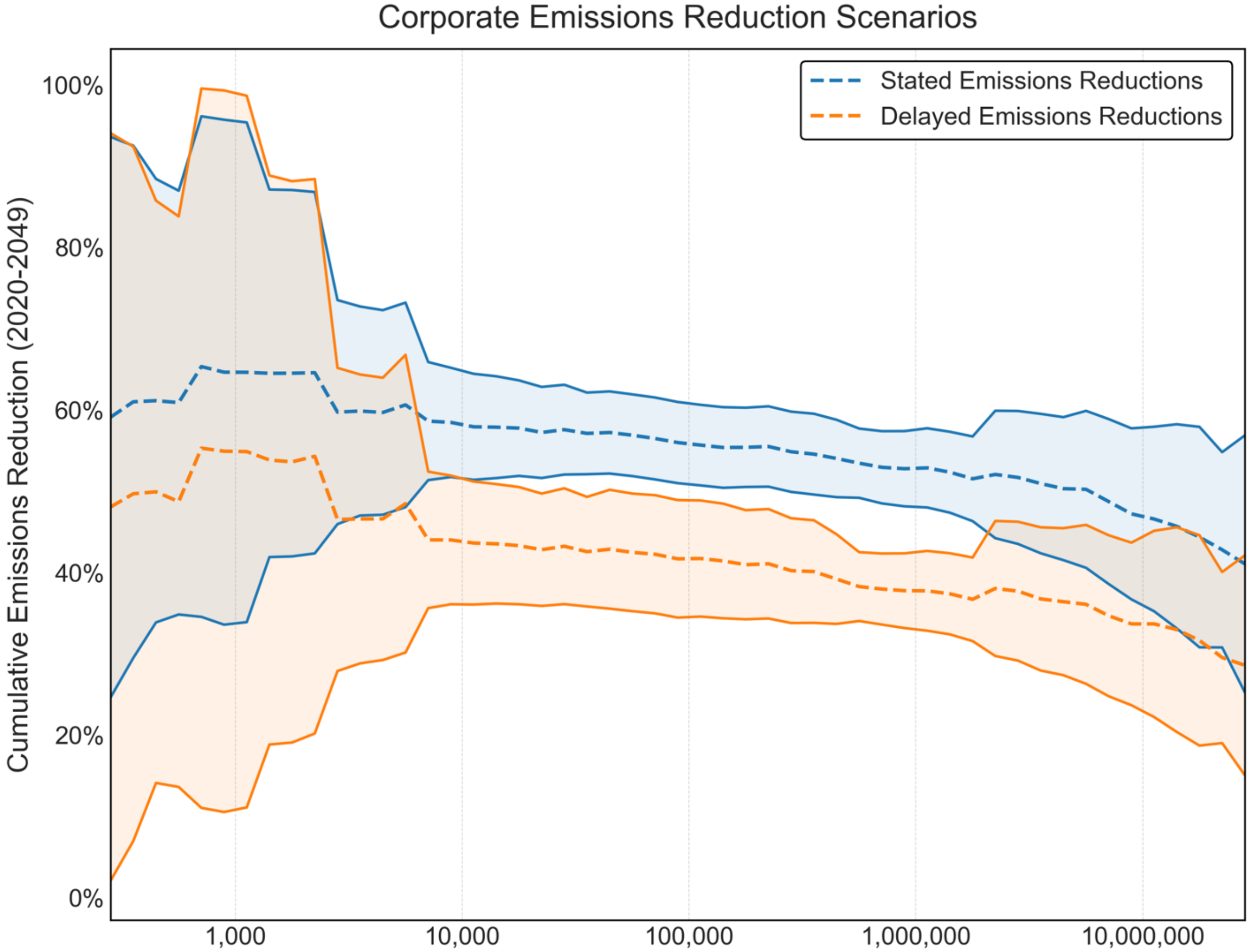

Figure 1: Percent reduction of cumulative emissions during the simulated period (2020-2049) by total reported emissions during the historical period (2020-2023) for all companies using both the Stated Emissions Reductions and Delayed Emissions Reductions scenarios. Companies are grouped into logarithmically spaced buckets, with 10 evenly spaced buckets per decade. The dashed line indicates the average, while the colored region displays the 95% confidence interval of the mean for each of these groups. A moving average with a width of 5 is applied to the resulting values. Note logarithmic x axis.

Emissions reduction plans vary by sector along with level of historical emissions, demonstrating how opportunity and ambition vary depending on the type of company (Figure 2). Real estate companies with the lowest historical emissions have the most aggressive emissions reduction plans, while large polluters in basic materials have the weakest emissions reduction plans, suggesting challenges for decarbonization in heavily polluting sectors. Utilities' stated emissions reduction targets result in over 50% cumulative emissions reductions for polluters of all sizes in this sector. This surprising result indicates that the analyzed utility companies are biased, as there are only 106 utility companies analyzed here, likely representing those which have the most ambitious reduction plans. Regardless, this suggests that substantial emissions reductions are at least plausible for utilities, even if currently this is only apparent in a subset of utility companies. This result is in contrast to materials companies, which even in this biased subset have low ambition, likely owing to the perceived difficulty or limited ambition for decarbonization in this sector. As materials companies are the most common major polluter in the dataset, this is a major driver of the lower decarbonization rates seen in larger polluters in Figure 1.

**Average Stated Emissions Reduction by Sector and Historical Emissions**

|  | 1K-10K (269) | 10K-100K (800) | 100K-1M (961) | 1M-10M (532) | 10M-100M (145) |
|---|---|---|---|---|---|
| Industrials (524) | 59% (30) | 58% (179) | 57% (216) | 58% (82) | 44% (17) |
| Consumer Cyclical (348) | 53% (32) | 60% (95) | 59% (157) | 58% (61) | 53% (2) |
| Technology (327) | 54% (47) | 55% (123) | 52% (112) | 55% (42) | 52% (3) |
| Basic Materials (303) | 65% (5) | 45% (21) | 48% (82) | 44% (130) | 34% (61) |
| Financial Services (235) | 57% (70) | 60% (117) | 56% (44) | 50% (3) | 45% (1) |
| Consumer Defensive (223) | 51% (9) | 50% (45) | 52% (111) | 52% (55) | 44% (3) |
| Healthcare (203) | 63% (20) | 57% (86) | 56% (81) | 57% (15) | None |
| Real Estate (165) | 75% (30) | 58% (73) | 50% (56) | 53% (6) | None |
| Energy (116) | 77% (1) | 56% (5) | 47% (26) | 47% (53) | 47% (28) |
| Utilities (106) | 66% (6) | 53% (9) | 53% (23) | 51% (42) | 54% (24) |
| Communication Services (102) | 61% (13) | 60% (29) | 68% (32) | 62% (25) | 68% (3) |

2020-2023 Annual $CO_2$ Emissions [tonnes]

Figure 2: Percent reduction of cumulative emissions following stated emissions reduction plans during the simulated period (2020-2049) by sector and annual reported emissions from 2020-2023. Numbers in parenthesis in rows and columns denote the total number of companies in each sector and reported emissions range respectively, while numbers in parenthesis in each cell denote the number of companies in each sector with a given level of historical emissions. Note that cell values in each row may not sum to sector value due to companies with annual reported emissions outside of the presented emissions levels.

To further analyze companies' emissions reduction plans, we calculate the annual emissions reduction rates associated with stated targets (Figure 3). The average (and median) company aims to reduce emissions at a rate of around 5% per year. Annual emissions reduction targets span a wide range, from less than 0.5% to over 22%, though the vast majority of targets are less than 10% per year. The prevalence of targets less than 5% per year suggests ample room for increased ambition in decarbonization, particularly in larger polluters and for companies in the materials sector as shown earlier. The median baseline year for emissions reductions is 2020, while the median target year is 2030, with target years as far in the future as 2060 (Supplement Fig. S2).

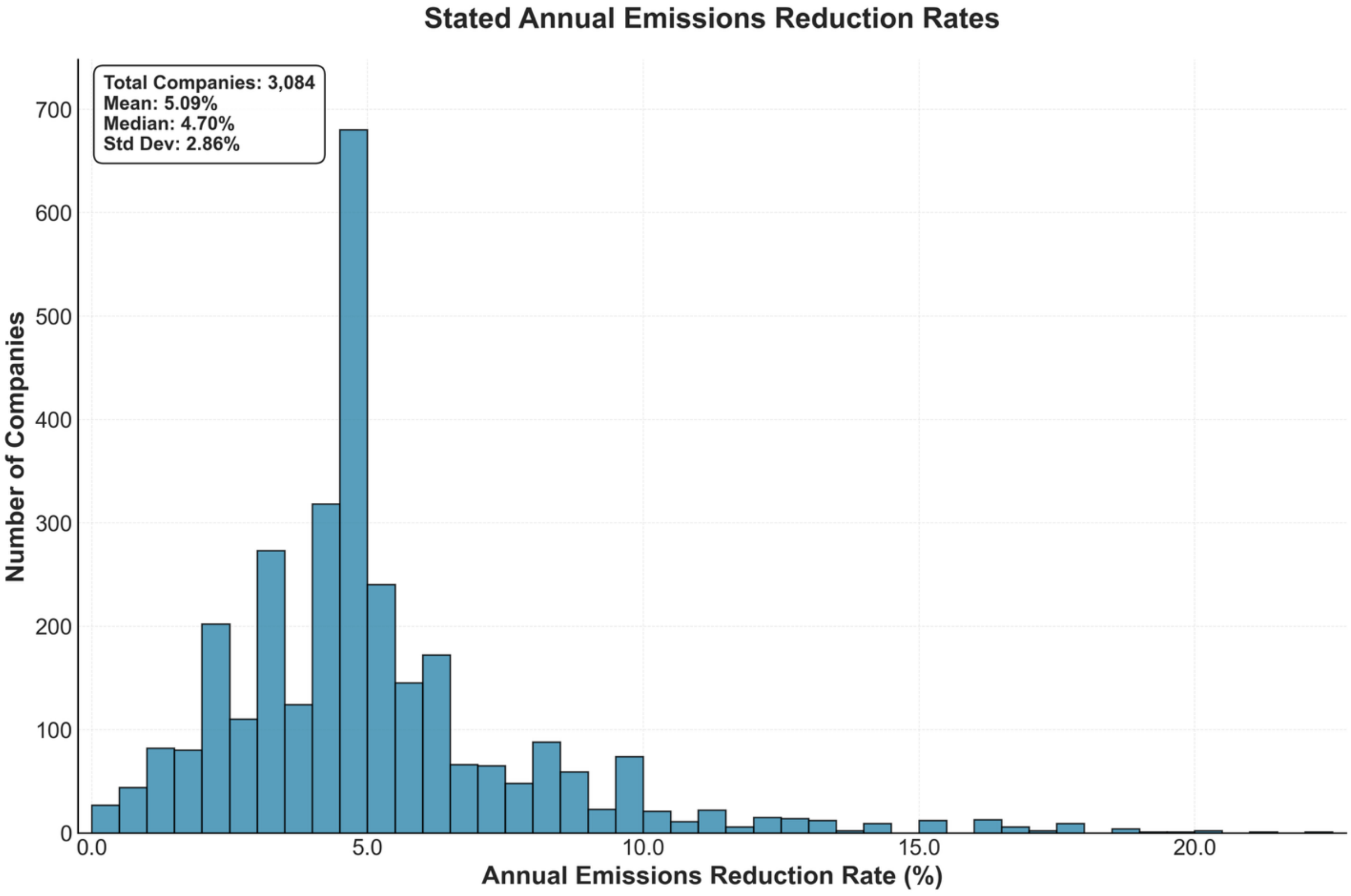

Figure 3: Distribution of annual emissions reduction percentage rates from companies' stated pledges. The annual percent reduction is calculated by dividing the reduction target by the number of years between the baseline and target year.

Next, we analyze the avoided warming that would follow implementation of emissions cuts. To calculate temperature increases from companies' individual emissions, we run FaIR simulations with and without a companies' emissions included, comparing the resulting warming. While global warming is often framed as a societal issue, the FaIR simulations indicate that implementation of emissions reductions pledges from even the smallest polluters produces quantifiable changes (Figure 4). As expected, the largest companies have the greatest potential to decrease warming, with the largest polluters (over 100,000,000 tonnes $CO_{2e}$ of annual emissions) capable of reducing warming by roughly 1/1,000 of a degree Celsius by following through with their planned emissions reductions. By comparison, the polluters with annual emissions between 1,000 and 10,000 tonnes $CO_{2e}$ are only capable of reducing warming by $10^{-8}$ degrees Celsius. Regardless, these FaIR simulations indicate that individual emissions reductions in the absence of collective action still produce calculable impacts on global warming, which we show in the next section result in meaningful decreases in heat-related deaths. Similar figures comparing the Delayed Emissions Reductions scenario to the other two scenarios are included in the Supplement.

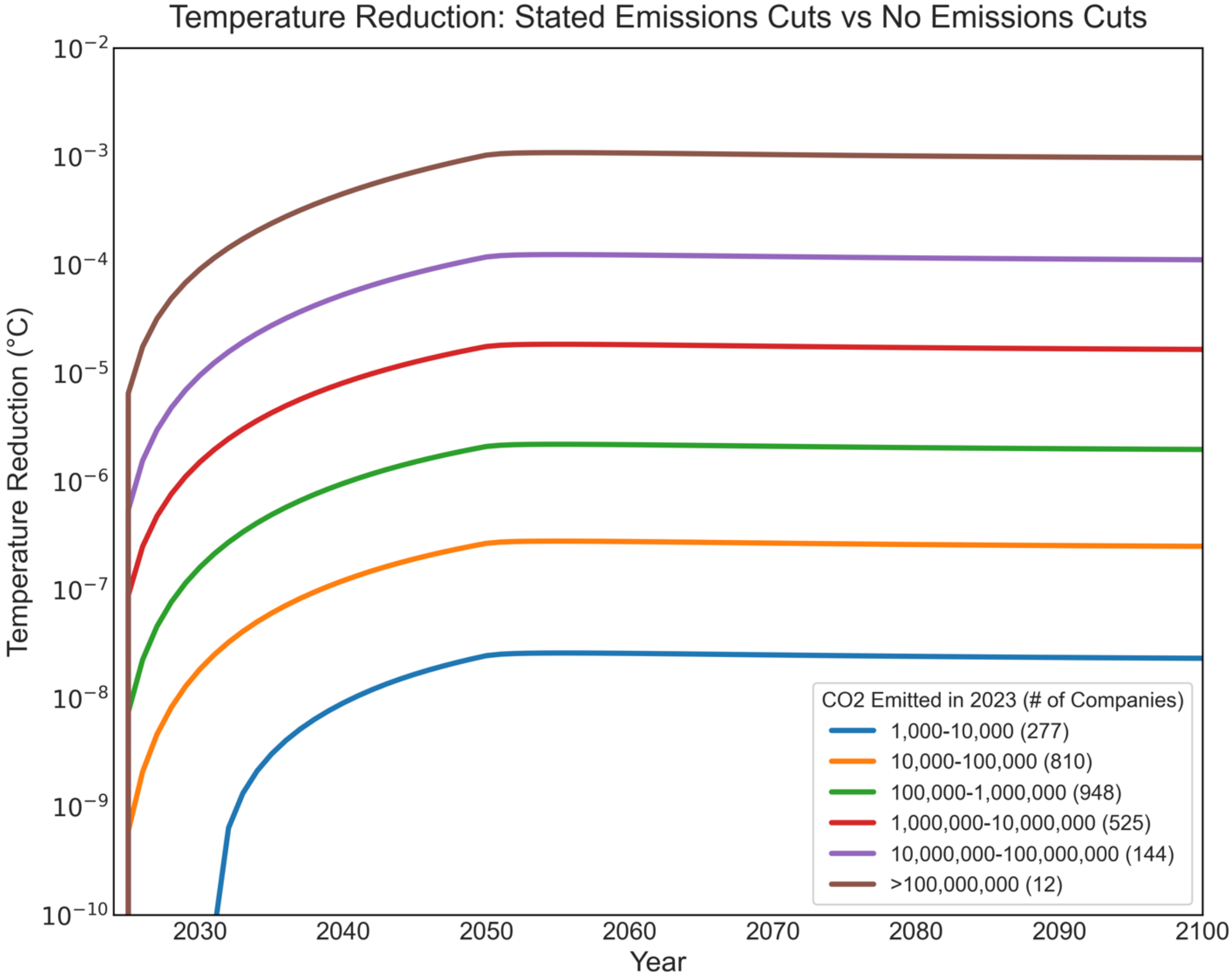

Figure 4: Average temperature reductions in each year resulting from emissions reduction pledges. Companies are grouped by 2023 $CO_{2e}$ emissions, with colored lines indicating the average temperature reduction for companies of a given amount of 2023 emissions. Note logarithmic y axis.

While these temperature impacts are quantifiable, it is difficult to envision the importance of $10^{-8}$ degrees Celsius of avoided warming. To contextualize these impacts, we utilize mortality damage functions from Wells et al. (2025), which are a globalized form of the country-level damage functions from Bressler et al. (2021) adapted from the observed relationships between heat and mortality in Gasparrini et al. (2017). These mortality damage functions use observed relationships between temperature and all-cause mortality to project potential heat and cold-related deaths from global warming. These death projections do not consider impacts from intensified weather, sea-level rise, co-benefits such as reduced air pollution, or secondary societal impacts such as resource scarcity. As such, these estimates likely represent a lower-bound for the life-saving potential of emissions reductions.

To calculate the potential lives saved from emissions reductions, we apply the mortality damage functions to temperature time series from the Stated Emissions Reductions and No Reductions scenarios and compare the results (Figure 5). Unsurprisingly, we find that larger polluters have the potential to save orders of magnitude more lives by following through with emissions reduction plans than do smaller polluters. However, we do find that even companies with annual emissions less than 1,000 tonnes $CO_{2e}$ have the potential to save at least one life by implementing planned emissions reductions. Less than 8% of companies were found to not have the potential to save at least one life. This results from the combination of low annual emissions and weak emissions reduction targets. Across all 3,084 companies, we find that implementation of pledges would result in over 4.4 million avoided temperature-related deaths by 2100. Similar figures for the Delayed Emissions Reductions scenario and estimates of uncertainty in the mortality calculations are included in the Supplement.

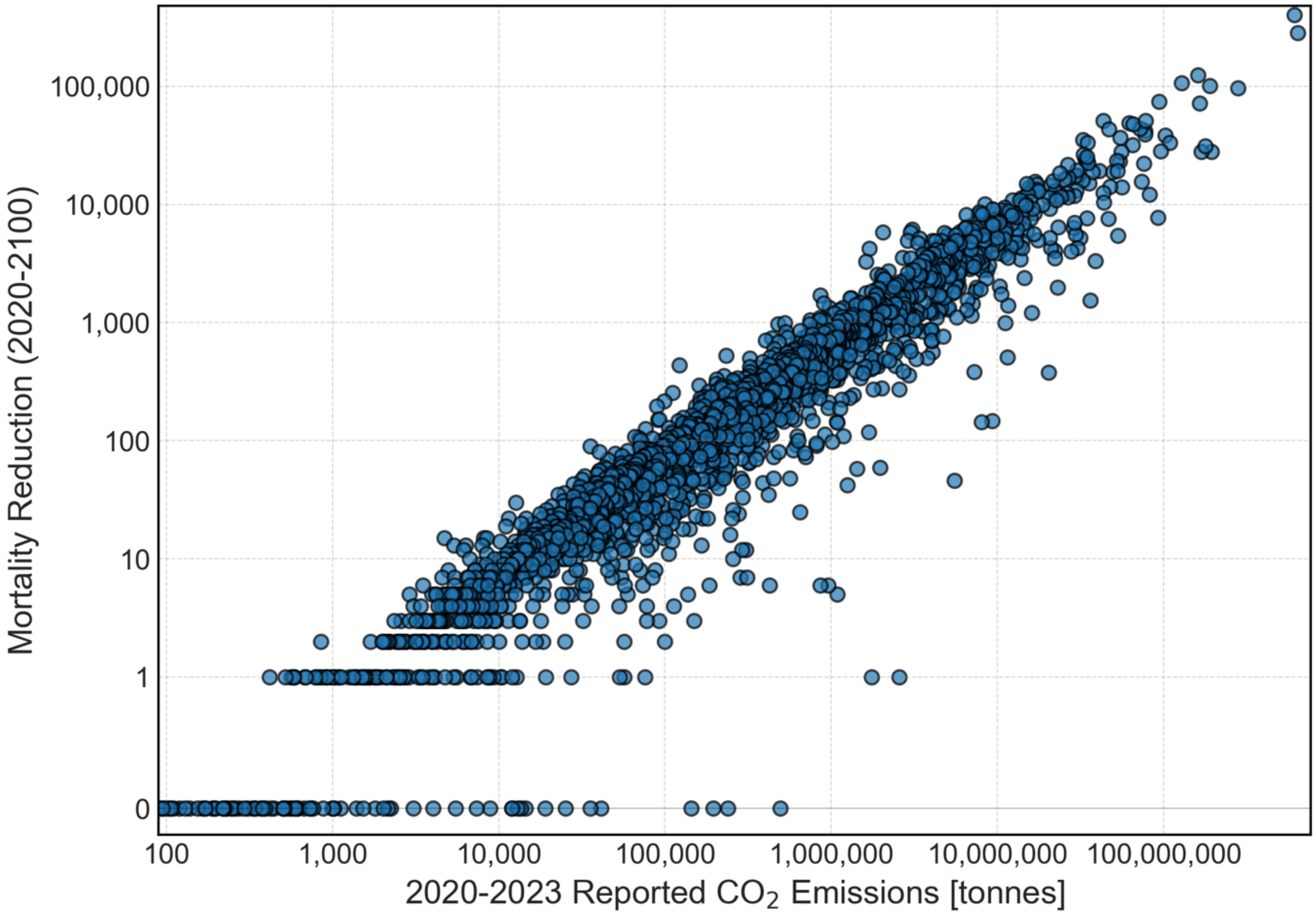

Figure 5: Total decrease in cumulative deaths by 2100 from pledged emissions reductions by annual average reported emissions from 2020-2023 for individual companies. Mortality reduction is calculated from the difference in the deaths calculated from the mortality damage functions for two scenarios: Pledged Emissions Reductions and No Emissions Reductions. Note logarithmic x and y axes, as well as rounding down to nearest integer.

Comparing by sector, we find that the highest percentage of companies with the potential to save at least one life by following through with their emissions reduction plans are in the basic materials sector, with the lowest percentage in the financial services sector (Figure 6). These trends are similar across higher mortality reduction thresholds, with the main difference being the above average 39-60% of energy and utility companies capable of preventing at least 1,000 temperature-related deaths despite these sectors being below average in percentage of companies preventing at least one temperature-related death. The potential for roughly half of companies in these three sectors to prevent at least 1,000 temperature-related deaths from their own planned actions shows the importance of these independent decarbonization efforts, demonstrating that even in the absence of coordinated, international action companies can make a substantial impact on outcomes. Additionally, in all sectors at least 80% of analyzed companies prevent at least one temperature-related death, suggesting that the potential for meaningful and quantifiable impact is present in most companies regardless of sector, even if the largest potential exists only in the three highest polluting sectors.

Percentage of Companies with Mortality Reductions by Sector: Stated Emissions Cuts versue No Cuts

| | ≥ 1 Death | ≥ 10 Deaths | ≥ 100 Deaths | ≥ 1,000 Deaths |
|---|---|---|---|---|
| Basic Materials (307) | 95-96% | 91-94% | 77-85% | 39-53% |
| Healthcare (206) | 94-96% | 78-87% | 32-46% | 1-6% |
| Consumer Defensive (225) | 95-96% | 85-88% | 54-68% | 12-20% |
| Industrials (531) | 94-95% | 82-87% | 40-54% | 9-17% |
| Consumer Cyclical (355) | 93-95% | 80-86% | 45-57% | 9-18% |
| Energy (119) | 93% | 91-92% | 82-87% | 49-60% |
| Utilities (107) | 89-92% | 82-84% | 70-75% | 39-50% |
| Technology (340) | 88-91% | 66-76% | 30-40% | 6-11% |
| Communication Services (110) | 85-88% | 70-74% | 41-54% | 13-23% |
| Real Estate (171) | 84-87% | 58-68% | 18-27% | 1-3% |
| Financial Services (263) | 82-86% | 48-57% | 8-14% | 0-1% |

Figure 6: Percentage of companies with at least 1, 10, 100, or 1,000 deaths by sector from stated emissions reduction plan. Values presented in each cell are the 5$^{th}$ and 95$^{th}$ percentile values from 10,000 Monte Carlo simulations varying the coefficients in the mortality damage function and the all-cause mortality estimates. Numbers in parenthesis in each row denote the number of companies in each sector.

Finally, we analyze four prescribed emissions reduction scenarios for the 3,084 companies to see the avoided damages resulting from various levels of reductions. We create four scenarios with reductions of 1%, 2%, 5%, and 10% per year respectively, and compare these scenarios to a no-reduction scenario. We find that companies with annual emissions over 1,000 tonnes $CO_{2e}$ have the potential to save at least one life by reducing emissions just 1% per year (Figure 5). Additionally, even some of the least polluting companies, with annual emissions less than 1,000 tonnes $CO_{2e}$, have the potential to save lives by reducing emissions at an annual rate of 5% or greater. Larger mortality reductions of 10, 100, or 1,000 lives are only possible for larger polluters, and become more probable with increasing decarbonization ambition. For the largest polluters, even minor reductions of 1% per year have large implications, indicating the continued importance of pushing for emissions cuts from the largest polluters. Similar to the emissions reduction pledge analysis, nearly all companies with pollution greater than 1,000 tonnes of $CO_{2e}$ annually have the potential to save lives by implementing single digit annual reduction rates. Taken together, these results suggest that even smaller companies have the potential to create meaningful impacts in the absence of collective action.

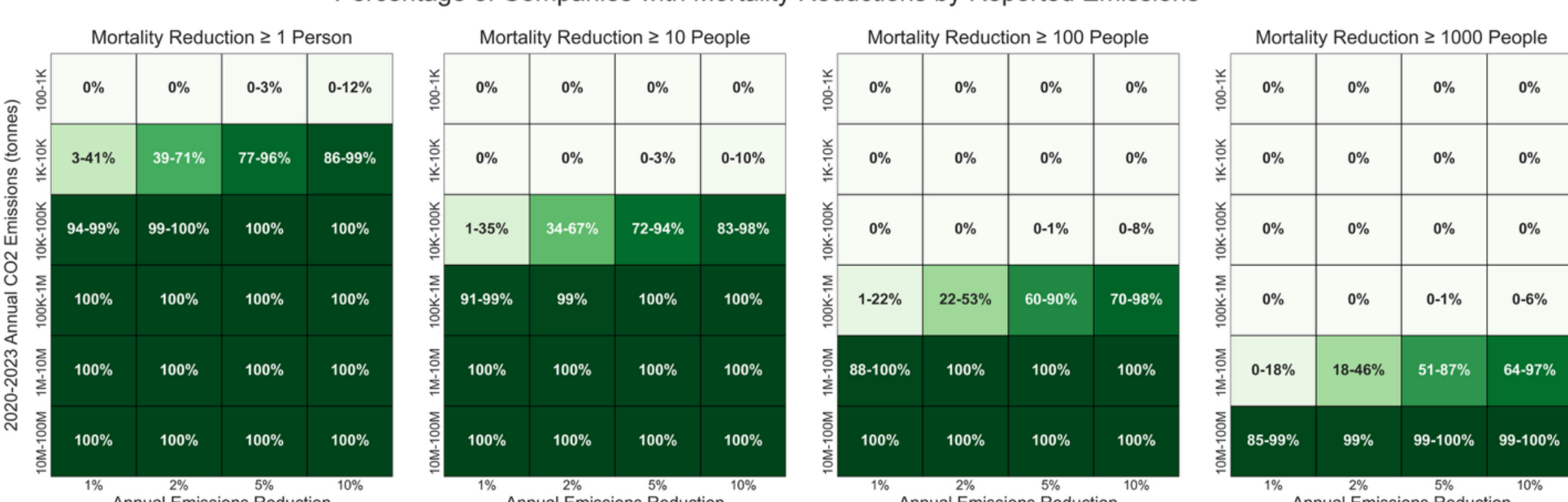

Figure 7: Percent of companies saving at least 1, 10, 100, or 1000 lives by reducing emissions by 1%, 2%, 5% or 10% per year, grouped by 2023 reported $CO_{2e}$ emissions. Values presented in each cell are the 5$^{th}$ and 95$^{th}$ percentiles from 10,000 Monte Carlo simulations computed by varying baseline all-cause mortality as well as mortality damage function coefficients.

## Summary and Discussion

We combined reported $CO_{2e}$ emissions and emissions reduction targets from Tracenable (Tracenable, 2025) to estimate the decrease in 2020-2049 $CO_{2e}$ emissions resulting from achievement of corporate sustainability targets. Using these emissions, we calculated the reduction in warming using a climate emulator and calculated the resulting mortality reduction using damage functions (Wells et al., 2025). Over 92% of the companies examined can save lives by following through with their emissions reduction plans. Furthermore, cuts from all 3,084 companies could save over 4.4 million lives.

Through this analysis, we present global warming as a problem that can be partially mitigated through action by individual companies, even in the absence of coordinated international efforts. Contrary to the typical representation of global warming as an issue that can only be impacted by global efforts, these results suggest that measurable improvements in outcome can occur from independent corporate actions. By highlighting the life-saving potential that even smaller companies possess, we hope to increase ambition in corporate sustainability and create new internal and external pressures to improve corporate sustainability efforts. In the absence of sufficient national and international regulation, voluntary corporate emissions reductions have the potential to meaningfully decrease warming and resulting impacts given adequate ambition, which we hope these visualizations will contribute to furthering.

There are a few key limitations of this analysis which present opportunities for building on this work. First, the mortality functions assume that the historical relationship between temperature and mortality is valid for 2020-2100, which is unlikely to be the case due to nonlinearities in human health responses to temperature as well as potential adaptation measures such as increases in air conditioning availability. The mortality functions are further limited by focusing solely on temperature-related mortality, neglecting impacts from chronic stressors, warming-fueled extreme weather, co-benefits, or second order effects. Last, and perhaps most significant, the analysis assumes constant 2023 Scope 1 and 2 emissions through 2049, which is unrealistically low for small companies that are aiming for rapid expansion, and also may overestimate baseline emissions from larger companies as carbon intensity decreases. An improved baseline scenario, or perhaps a probabilistic modeling of baseline emissions would further elucidate the potential of individual companies' sustainability plans. Taken together, these limitations likely cause the company-specific mortality analysis to underestimate life-saving potential overall, largely due to the sole focus on temperature-related mortality, but may overestimate certain companies.

The methods presented can also be adapted to custom emissions reduction scenarios, allowing for bespoke analyses of the life-saving potential of specific decarbonization efforts. These analyses could be used to buttress targeted internal and external sustainability campaigns by sustainability managers, consultants, activists and regulators. Additionally, the methods here present a base from which further impacts can be assessed. The warming levels calculated from the climate emulator allow for the calculation of damages from a variety of impacts, either using damage functions as demonstrated here or by modeling physical impacts through a warming-levels based analysis. For these analyses, we recommend using aggregated statistics, as the impacts from the minor temperature differences are unlikely to be visible otherwise. Further research could involve the development of new damage functions or the expansion of this topic through novel metrics of aggregated climate impacts.

**Acknowledgements**

The authors thank Tracenable (tracenable.com) for providing reported and pledged emissions at a reduced rate.

## Appendix

### Text S1: Distribution of analyzed companies by sector and country

Top five sector-countries are United States industrials, United States technology, Taiwan technology, Japan industrials, and United States communication services. Companies included in this analysis must have reported emissions and a stated emissions reduction plan in the Tracenable database, causing the mix of companies to be an imperfect representation of all global companies.

Percentage of Total Companies by Sector and Country

| | Basic Materials (10.7%) | Communication Services (3.6%) | Consumer Cyclical (12.3%) | Consumer Defensive (7.7%) | Energy (4.0%) | Financial Services (8.8%) | Healthcare (7.3%) | Industrials (18.6%) | Real Estate (6.1%) | Technology (12.0%) | Utilities (3.6%) |
|---|---|---|---|---|---|---|---|---|---|---|---|
| United States (20.1%) | 1.4% | 0.6% | 3.1% | 1.6% | 1.1% | 1.3% | 1.8% | 3.5% | 1.6% | 3.4% | 0.7% |
| Japan (10.9%) | 1.0% | 0.2% | 1.8% | 0.9% | 0.1% | 0.7% | 0.8% | 3.3% | 0.5% | 1.4% | 0.2% |
| Taiwan (7.5%) | 1.0% | 0.1% | 0.7% | 0.1% | 0.0% | 1.0% | 0.1% | 1.0% | 0.1% | 3.4% | 0.0% |
| China (Mainland) (5.6%) | 1.0% | 0.2% | 0.6% | 0.4% | 0.2% | 0.2% | 1.0% | 0.5% | 0.4% | 1.0% | 0.1% |
| United Kingdom (5.3%) | 0.4% | 0.4% | 1.0% | 0.5% | 0.1% | 0.7% | 0.3% | 1.1% | 0.3% | 0.2% | 0.3% |
| India (3.4%) | 0.7% | 0.1% | 0.3% | 0.3% | 0.1% | 0.3% | 0.5% | 0.5% | 0.1% | 0.4% | 0.1% |
| South Korea (3.1%) | 0.3% | 0.2% | 0.4% | 0.3% | 0.1% | 0.5% | 0.2% | 0.9% | 0.0% | 0.2% | 0.0% |
| Canada (3.1%) | 0.6% | 0.1% | 0.3% | 0.3% | 0.5% | 0.4% | 0.0% | 0.4% | 0.2% | 0.1% | 0.2% |
| Germany (3.0%) | 0.4% | 0.1% | 0.7% | 0.1% | 0.0% | 0.2% | 0.4% | 0.7% | 0.1% | 0.1% | 0.2% |
| France (2.9%) | 0.2% | 0.2% | 0.5% | 0.3% | 0.1% | 0.2% | 0.2% | 0.7% | 0.2% | 0.3% | 0.0% |
| Australia (2.4%) | 0.5% | 0.1% | 0.2% | 0.1% | 0.2% | 0.4% | 0.2% | 0.4% | 0.2% | 0.0% | 0.1% |
| Hong Kong (2.4%) | 0.1% | 0.1% | 0.1% | 0.2% | 0.0% | 0.1% | 0.1% | 0.6% | 0.7% | 0.1% | 0.3% |
| Switzerland (2.3%) | 0.2% | 0.0% | 0.4% | 0.2% | 0.0% | 0.4% | 0.3% | 0.5% | 0.1% | 0.2% | 0.0% |
| Sweden (2.2%) | 0.2% | 0.2% | 0.2% | 0.2% | 0.0% | 0.1% | 0.2% | 0.7% | 0.3% | 0.1% | 0.0% |
| Thailand (2.1%) | 0.2% | 0.1% | 0.2% | 0.2% | 0.2% | 0.2% | 0.1% | 0.2% | 0.3% | 0.1% | 0.3% |
| South Africa (1.4%) | 0.4% | 0.1% | 0.1% | 0.1% | 0.1% | 0.1% | 0.1% | 0.1% | 0.2% | 0.1% | 0.0% |
| Brazil (1.4%) | 0.4% | 0.0% | 0.2% | 0.3% | 0.0% | 0.1% | 0.0% | 0.3% | 0.0% | 0.0% | 0.1% |
| Turkey (1.3%) | 0.2% | 0.1% | 0.2% | 0.1% | 0.0% | 0.3% | 0.0% | 0.3% | 0.0% | 0.0% | 0.1% |
| Denmark (1.3%) | 0.1% | 0.0% | 0.1% | 0.1% | 0.0% | 0.2% | 0.3% | 0.4% | 0.0% | 0.1% | 0.0% |
| Netherlands (1.2%) | 0.1% | 0.1% | 0.0% | 0.1% | 0.1% | 0.1% | 0.1% | 0.3% | 0.1% | 0.2% | 0.0% |
| Malaysia (1.1%) | 0.1% | 0.0% | 0.0% | 0.2% | 0.1% | 0.2% | 0.1% | 0.2% | 0.1% | 0.1% | 0.0% |
| Spain (1.0%) | 0.0% | 0.0% | 0.2% | 0.0% | 0.0% | 0.1% | 0.1% | 0.2% | 0.1% | 0.1% | 0.2% |
| Finland (1.0%) | 0.1% | 0.0% | 0.2% | 0.2% | 0.0% | 0.1% | 0.0% | 0.3% | 0.0% | 0.1% | 0.0% |
| Italy (1.0%) | 0.1% | 0.0% | 0.2% | 0.0% | 0.1% | 0.1% | 0.0% | 0.3% | 0.0% | 0.0% | 0.2% |
| Mexico (0.7%) | 0.1% | 0.1% | 0.0% | 0.2% | 0.0% | 0.0% | 0.0% | 0.2% | 0.1% | 0.0% | 0.0% |
| Singapore (0.7%) | 0.0% | 0.1% | 0.0% | 0.1% | 0.0% | 0.0% | 0.0% | 0.2% | 0.2% | 0.1% | 0.0% |
| Ireland (0.7%) | 0.1% | 0.0% | 0.1% | 0.1% | 0.0% | 0.0% | 0.1% | 0.2% | 0.0% | 0.1% | 0.0% |
| Indonesia (0.7%) | 0.2% | 0.1% | 0.0% | 0.1% | 0.2% | 0.0% | 0.1% | 0.0% | 0.0% | 0.0% | 0.0% |
| Belgium (0.6%) | 0.1% | 0.0% | 0.0% | 0.1% | 0.0% | 0.0% | 0.1% | 0.2% | 0.1% | 0.0% | 0.0% |
| Poland (0.5%) | 0.1% | 0.1% | 0.1% | 0.0% | 0.1% | 0.1% | 0.0% | 0.0% | 0.0% | 0.0% | 0.0% |
| Austria (0.5%) | 0.0% | 0.0% | 0.1% | 0.0% | 0.0% | 0.2% | 0.0% | 0.2% | 0.0% | 0.0% | 0.0% |
| Chile (0.4%) | 0.1% | 0.0% | 0.0% | 0.0% | 0.0% | 0.1% | 0.0% | 0.1% | 0.1% | 0.0% | 0.0% |
| New Zealand (0.4%) | 0.0% | 0.1% | 0.0% | 0.0% | 0.0% | 0.0% | 0.1% | 0.1% | 0.0% | 0.0% | 0.1% |
| Norway (0.4%) | 0.1% | 0.0% | 0.1% | 0.1% | 0.1% | 0.0% | 0.0% | 0.0% | 0.0% | 0.0% | 0.0% |
| Greece (0.3%) | 0.0% | 0.0% | 0.0% | 0.0% | 0.1% | 0.1% | 0.0% | 0.0% | 0.0% | 0.0% | 0.1% |
| Luxembourg (0.3%) | 0.0% | 0.1% | 0.1% | 0.1% | 0.0% | 0.0% | 0.0% | 0.0% | 0.0% | 0.0% | 0.0% |
| Colombia (0.3%) | 0.0% | 0.0% | 0.0% | 0.0% | 0.1% | 0.1% | 0.0% | 0.0% | 0.0% | 0.0% | 0.1% |
| Saudi Arabia (0.3%) | 0.1% | 0.0% | 0.0% | 0.0% | 0.1% | 0.0% | 0.0% | 0.0% | 0.0% | 0.0% | 0.1% |
| Portugal (0.2%) | 0.0% | 0.0% | 0.0% | 0.1% | 0.0% | 0.0% | 0.0% | 0.0% | 0.0% | 0.0% | 0.1% |
| Russia (0.2%) | 0.1% | 0.0% | 0.0% | 0.0% | 0.1% | 0.0% | 0.0% | 0.0% | 0.0% | 0.0% | 0.0% |
| Israel (0.1%) | 0.0% | 0.0% | 0.0% | 0.0% | 0.0% | 0.0% | 0.0% | 0.0% | 0.0% | 0.1% | 0.0% |
| Qatar (0.1%) | 0.0% | 0.0% | 0.0% | 0.0% | 0.0% | 0.1% | 0.0% | 0.0% | 0.0% | 0.0% | 0.0% |
| Bermuda (0.1%) | 0.0% | 0.0% | 0.0% | 0.0% | 0.0% | 0.1% | 0.0% | 0.0% | 0.0% | 0.0% | 0.0% |
| Macau (0.1%) | 0.0% | 0.0% | 0.1% | 0.0% | 0.0% | 0.0% | 0.0% | 0.0% | 0.0% | 0.0% | 0.0% |
| Romania (0.1%) | 0.0% | 0.0% | 0.0% | 0.0% | 0.1% | 0.0% | 0.0% | 0.0% | 0.0% | 0.0% | 0.0% |
| United Arab Emirates (0.0%) | 0.0% | 0.0% | 0.0% | 0.0% | 0.0% | 0.0% | 0.0% | 0.0% | 0.0% | 0.0% | 0.0% |
| Argentina (0.0%) | 0.0% | 0.0% | 0.0% | 0.0% | 0.0% | 0.0% | 0.0% | 0.0% | 0.0% | 0.0% | 0.0% |
| Peru (0.0%) | 0.0% | 0.0% | 0.0% | 0.0% | 0.0% | 0.0% | 0.0% | 0.0% | 0.0% | 0.0% | 0.0% |
| Panama (0.0%) | 0.0% | 0.0% | 0.0% | 0.0% | 0.0% | 0.0% | 0.0% | 0.0% | 0.0% | 0.0% | 0.0% |
| Monaco (0.0%) | 0.0% | 0.0% | 0.0% | 0.0% | 0.0% | 0.0% | 0.0% | 0.0% | 0.0% | 0.0% | 0.0% |
| Lithuania (0.0%) | 0.0% | 0.0% | 0.0% | 0.0% | 0.0% | 0.0% | 0.0% | 0.0% | 0.0% | 0.0% | 0.0% |
| Kazakhstan (0.0%) | 0.0% | 0.0% | 0.0% | 0.0% | 0.0% | 0.0% | 0.0% | 0.0% | 0.0% | 0.0% | 0.0% |
| Jersey (0.0%) | 0.0% | 0.0% | 0.0% | 0.0% | 0.0% | 0.0% | 0.0% | 0.0% | 0.0% | 0.0% | 0.0% |
| Isle of Man (0.0%) | 0.0% | 0.0% | 0.0% | 0.0% | 0.0% | 0.0% | 0.0% | 0.0% | 0.0% | 0.0% | 0.0% |
| Iceland (0.0%) | 0.0% | 0.0% | 0.0% | 0.0% | 0.0% | 0.0% | 0.0% | 0.0% | 0.0% | 0.0% | 0.0% |
| Hungary (0.0%) | 0.0% | 0.0% | 0.0% | 0.0% | 0.0% | 0.0% | 0.0% | 0.0% | 0.0% | 0.0% | 0.0% |
| Faroe Islands (0.0%) | 0.0% | 0.0% | 0.0% | 0.0% | 0.0% | 0.0% | 0.0% | 0.0% | 0.0% | 0.0% | 0.0% |
| Czechia (0.0%) | 0.0% | 0.0% | 0.0% | 0.0% | 0.0% | 0.0% | 0.0% | 0.0% | 0.0% | 0.0% | 0.0% |
| Cayman Islands (0.0%) | 0.0% | 0.0% | 0.0% | 0.0% | 0.0% | 0.0% | 0.0% | 0.0% | 0.0% | 0.0% | 0.0% |
| Kuwait (0.0%) | 0.0% | 0.0% | 0.0% | 0.0% | 0.0% | 0.0% | 0.0% | 0.0% | 0.0% | 0.0% | 0.0% |

Fig. S1: Percentage of analyzed companies by country (row) and sector (column), with cell value equal to the percentage by sector-country.

**Text S2: Distribution of baseline and target years**

Baseline years are typically between 2015 and 2025, with a few outliers as far in the past as 1990 and as far in the future as 2040. Target years generally vary more, and are most common in 2025, 2030, and 2035, with a few as far in the future as 2050 or 2060.

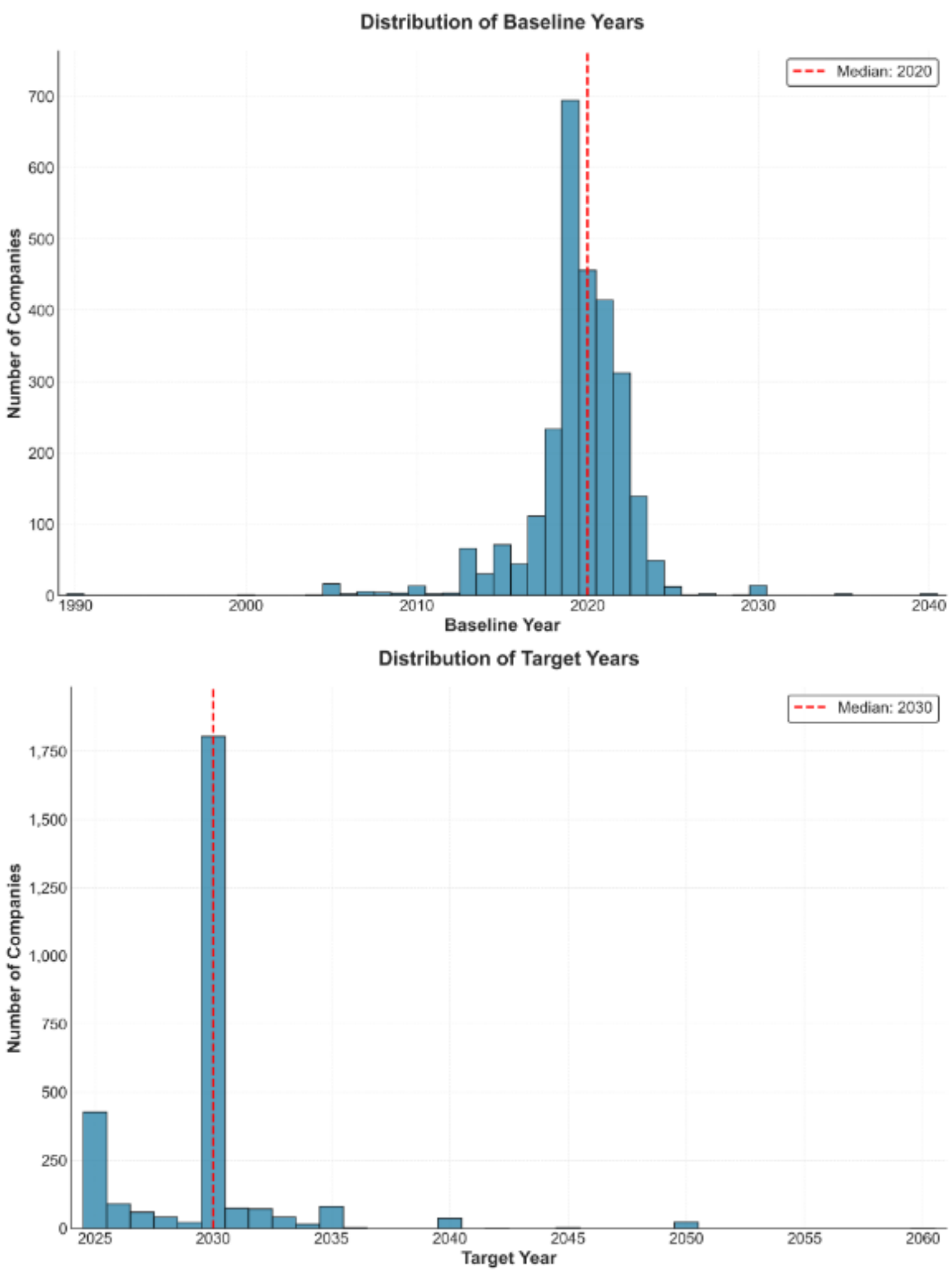

Fig. S2: Distribution of baseline and target years in analyzed companies' emissions reduction plans

### Text S3: Temperature changes for delayed emissions reductions scenario

Similar to the stated reductions scenario, a 5-year delay in reaching emissions reduction targets results in quantifiable warming reductions. The largest polluters reduce global temperatures by around 1/1000 of a degree Celsius, while the smallest polluters reduce warming by orders of magnitude less.

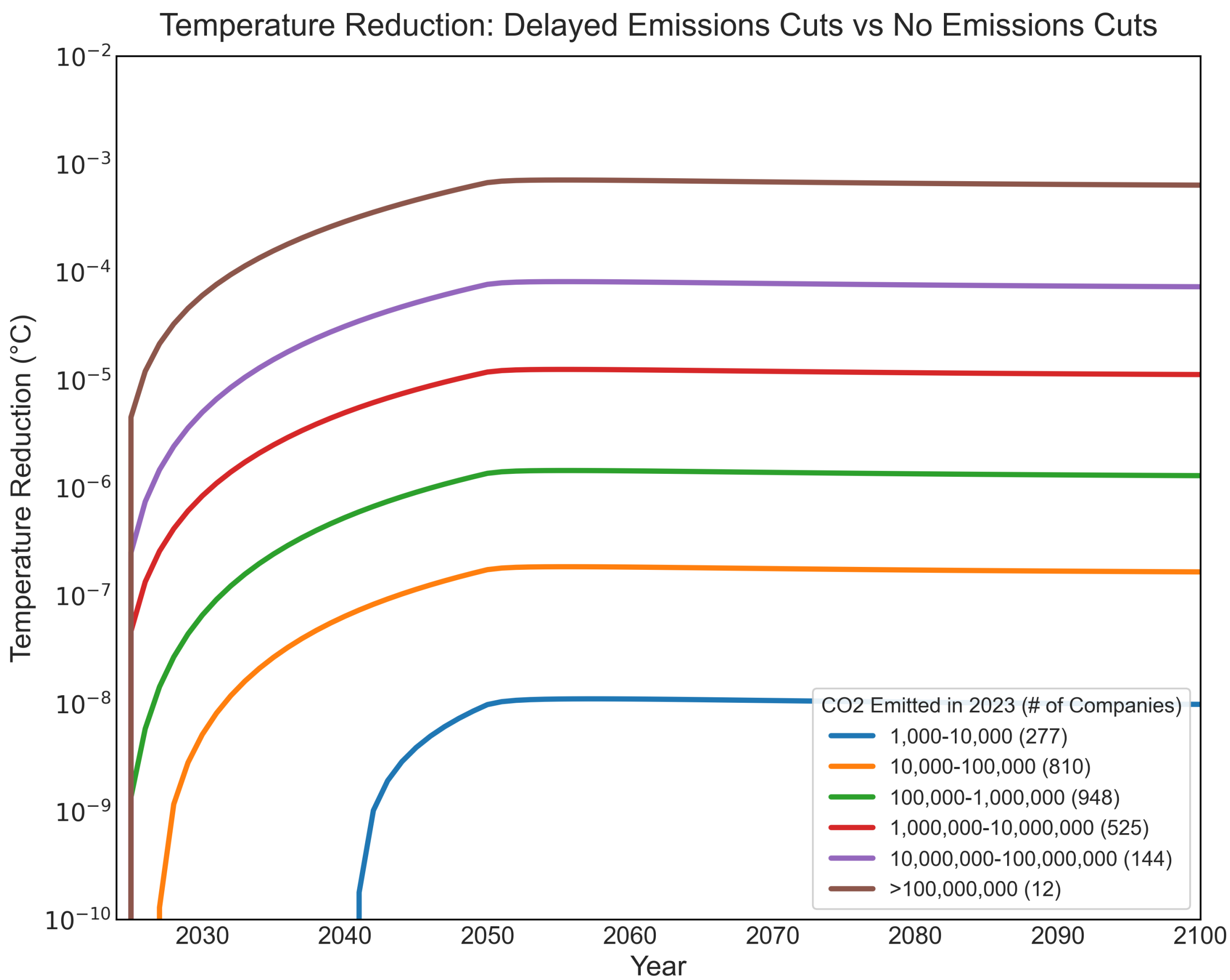

Fig. S3: Average temperature reductions in each year resulting from 5-year delayed emissions reductions scenario. Companies are grouped by 2023 $CO_{2e}$ emissions, with colored lines indicating the average temperature reduction for companies of a given amount of 2023 emissions. Note logarithmic y axis.

**Text S4: Mortality reduction from 5-year delayed emissions reductions scenario**

In the delayed emissions reduction scenario, the total lives saved across all 3,084 companies is reduced to around 2.9 million. The number of companies that do not save at least one life is increased to 131, demonstrating that even a delayed implementation of emissions cuts is impactful.

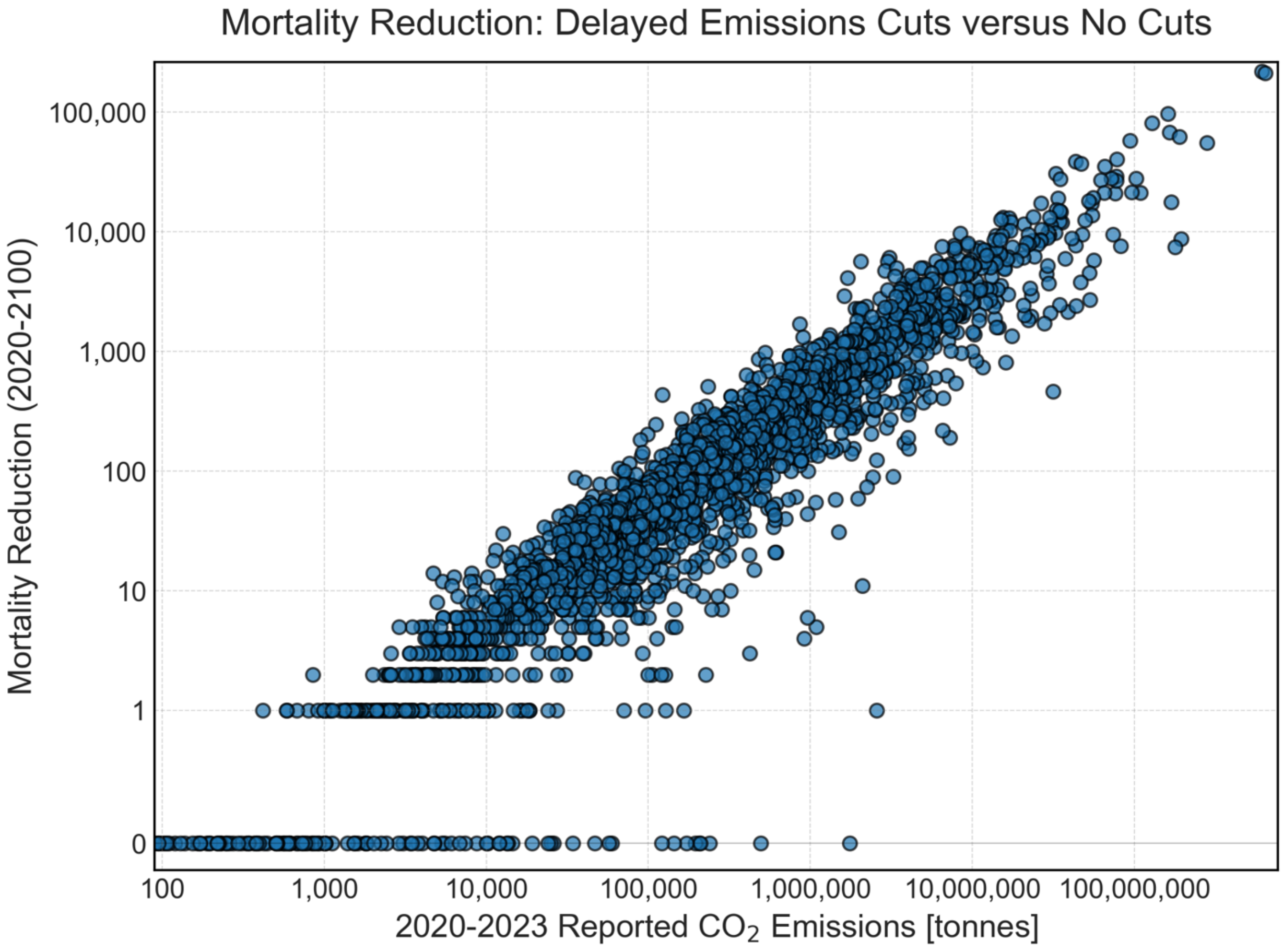

Fig. S4: Total decrease in cumulative deaths by 2100 from 5-year delayed emissions reductions scenario by annual average reported emissions from 2020-2023 for individual companies. Mortality reduction is calculated from the difference in the deaths calculated from the mortality damage functions for two scenarios: pledged emissions cuts and no emissions cuts. Note logarithmic x and y axes, as well as rounding down to nearest integer.

## Text S5: Uncertainty range of mortality reduction from emissions reductions

Mortality reduction at the 5th and 95th percentile was calculated from 10,000 Monte Carlo simulations varying the mortality damage function coefficients within the range from Wells et al. (2025) and varying the all-cause mortality values within the range from the United Nations World Population Prospects (2024). This does not account for uncertainty due to climate modeling in FaIR, global or individual corporate emissions, or implementation of corporate emissions reduction plans. Uncertainty from all-cause mortality and damage function coefficients is roughly 25% of average mortality projections, suggesting that the mortality values for individual companies should be treated as rough estimates.

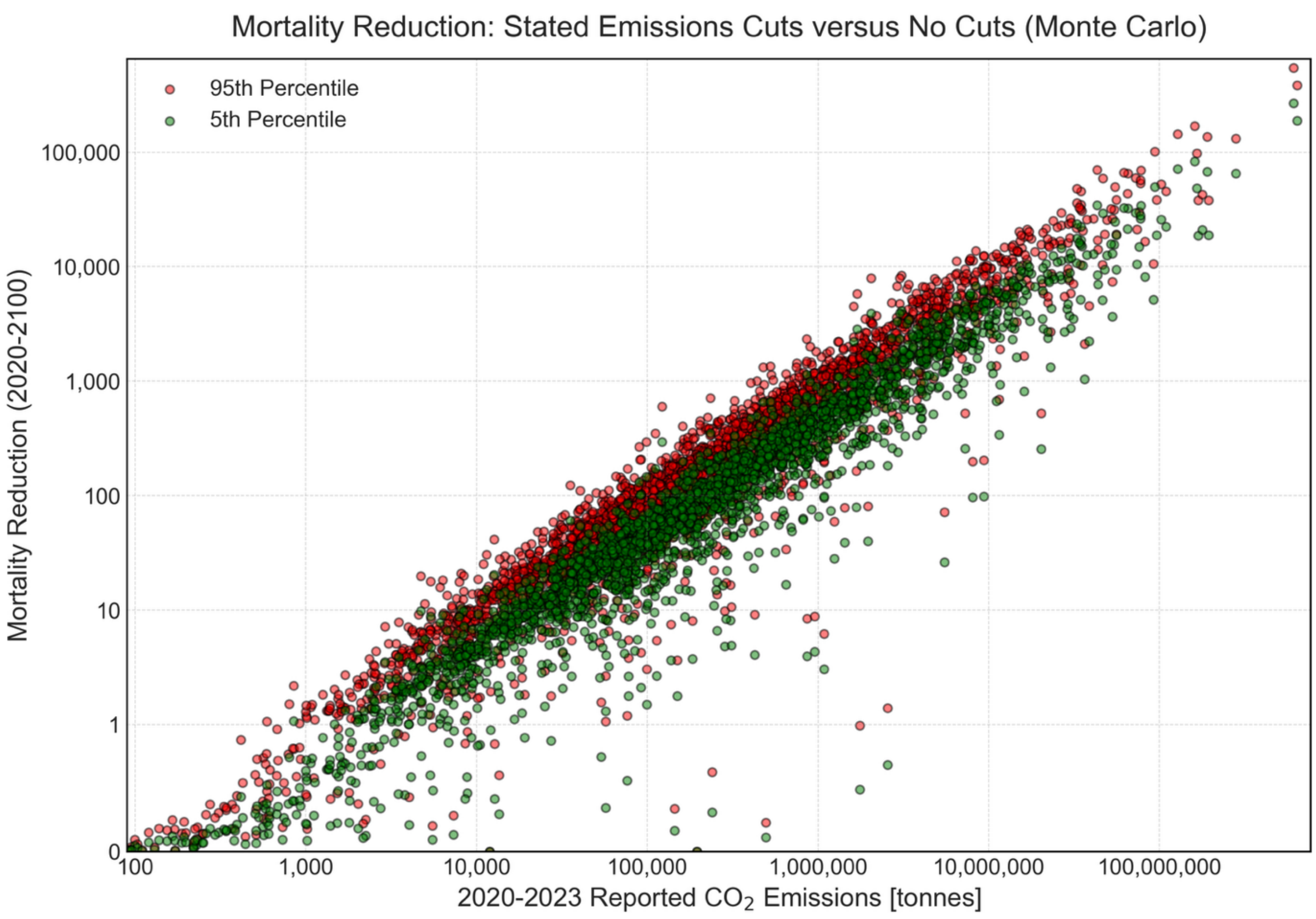

Fig. S5: 5th and 95th percentile of mortality reduction between stated emissions reduction plan and no reduction scenarios from 10,000 Monte Carlo simulations.

## Text S6: Percentage of companies reducing at least 1, 10, 100, and 1,000 deaths by country

Of the countries that are well represented in the dataset, proportions of companies capable of saving more than 1, 10, 100, 1,000 lives through their stated emissions reduction plans are fairly similar. The largest outliers are Saudi Arabia, with a large proportion of companies capable of saving more than 1,000 lives, likely due to the prevalence of carbon majors, and Hong Kong with a lower percentage of companies at all thresholds, which may be due to the relatively large number of analyzed companies in real estate.

Percentage of Companies with Mortality Reductions by Country:
Stated Emissions Cuts versus No Cuts

|  | ≥ 1 Death (93.8%) | ≥ 10 Deaths (80.6%) | ≥ 100 Deaths (48.1%) | ≥ 1,000 Deaths (17.4%) |
|---|---|---|---|---|
| Spain (1.2%) | 100.0% | 85.3% | 50.0% | 17.6% |
| Singapore (0.8%) | 100.0% | 81.8% | 50.0% | 13.6% |
| Austria (0.6%) | 100.0% | 87.5% | 56.2% | 25.0% |
| Chile (0.4%) | 100.0% | 83.3% | 50.0% | 25.0% |
| Saudi Arabia (0.4%) | 100.0% | 100.0% | 90.9% | 81.8% |
| Colombia (0.4%) | 100.0% | 80.0% | 70.0% | 20.0% |
| Greece (0.3%) | 100.0% | 88.9% | 55.6% | 44.4% |
| Philippines (0.3%) | 100.0% | 100.0% | 66.7% | 11.1% |
| Russia (0.3%) | 100.0% | 100.0% | 88.9% | 77.8% |
| United Arab Emirates (0.3%) | 100.0% | 88.9% | 66.7% | 44.4% |
| Bermuda (0.3%) | 100.0% | 62.5% | 12.5% | 12.5% |
| Portugal (0.3%) | 100.0% | 100.0% | 87.5% | 25.0% |
| Peru (0.1%) | 100.0% | 75.0% | 50.0% | |
| Cayman Islands (0.1%) | 100.0% | 100.0% | 50.0% | |
| Czechia (0.1%) | 100.0% | 50.0% | 50.0% | 50.0% |
| Hungary (0.1%) | 100.0% | 100.0% | 50.0% | 50.0% |
| Macau (0.1%) | 100.0% | 100.0% | 50.0% | |
| Qatar (0.1%) | 100.0% | 50.0% | | |
| Romania (0.1%) | 100.0% | 100.0% | 100.0% | 50.0% |
| Vietnam (0.1%) | 100.0% | 100.0% | 50.0% | 50.0% |
| Bangladesh (0.0%) | 100.0% | 100.0% | | |
| Croatia (0.0%) | 100.0% | 100.0% | | |
| Egypt (0.0%) | 100.0% | 100.0% | | |
| Faroe Islands (0.0%) | 100.0% | 100.0% | 100.0% | |
| Isle of Man (0.0%) | 100.0% | | | |
| Kenya (0.0%) | 100.0% | 100.0% | | |
| Kuwait (0.0%) | 100.0% | 100.0% | 100.0% | |
| Malta (0.0%) | 100.0% | 100.0% | | |
| Monaco (0.0%) | 100.0% | | | |
| Nigeria (0.0%) | 100.0% | 100.0% | | |
| Panama (0.0%) | 100.0% | 100.0% | 100.0% | 100.0% |
| Slovenia (0.0%) | 100.0% | 100.0% | | |
| Sri Lanka (0.0%) | 100.0% | | | |
| Germany (3.1%) | 98.9% | 87.5% | 56.8% | 25.0% |
| China (Mainland) (5.5%) | 98.1% | 87.0% | 59.7% | 26.0% |
| Italy (1.2%) | 97.0% | 90.9% | 54.5% | 18.2% |
| France (3.3%) | 96.7% | 81.5% | 51.1% | 19.6% |
| Canada (3.1%) | 96.6% | 86.2% | 60.9% | 26.4% |
| Japan (11.3%) | 96.2% | 89.9% | 56.6% | 15.5% |
| United States (20.3%) | 96.1% | 86.6% | 53.6% | 19.7% |
| India (3.4%) | 95.8% | 83.2% | 40.0% | 15.8% |
| Indonesia (0.8%) | 95.7% | 91.3% | 78.3% | 30.4% |
| Thailand (2.2%) | 95.2% | 74.6% | 47.6% | 25.4% |
| Brazil (1.5%) | 95.1% | 73.2% | 46.3% | 17.1% |
| Ireland (0.7%) | 95.0% | 85.0% | 65.0% | 15.0% |
| South Africa (1.3%) | 94.4% | 83.3% | 61.1% | 22.2% |
| Turkey (1.3%) | 94.4% | 83.3% | 52.8% | 30.6% |
| South Korea (3.1%) | 94.3% | 84.1% | 46.6% | 20.5% |
| Norway (0.6%) | 93.8% | 75.0% | 50.0% | 31.2% |
| Poland (0.5%) | 93.3% | 73.3% | 33.3% | 13.3% |
| United Kingdom (5.4%) | 92.8% | 70.4% | 33.6% | 9.2% |
| Denmark (1.4%) | 92.5% | 67.5% | 22.5% | 5.0% |
| Luxembourg (0.4%) | 91.7% | 91.7% | 33.3% | |
| Netherlands (1.2%) | 91.4% | 77.1% | 48.6% | 17.1% |
| Mexico (0.8%) | 91.3% | 78.3% | 47.8% | 13.0% |
| Switzerland (2.3%) | 90.8% | 67.7% | 33.8% | 3.1% |
| Sweden (2.1%) | 88.1% | 66.1% | 25.4% | 1.7% |
| Taiwan (7.5%) | 88.0% | 65.6% | 24.9% | 6.2% |
| Malaysia (1.1%) | 86.7% | 60.0% | 23.3% | 10.0% |
| Australia (2.3%) | 86.2% | 76.9% | 52.3% | 15.4% |
| Belgium (0.7%) | 84.2% | 68.4% | 36.8% | 15.8% |
| Israel (0.2%) | 83.3% | 66.7% | 33.3% | 16.7% |
| Hong Kong (2.6%) | 80.6% | 75.0% | 58.3% | 16.7% |
| New Zealand (0.4%) | 75.0% | 50.0% | 25.0% | 8.3% |
| Finland (1.2%) | 72.7% | 63.6% | 24.2% | 6.1% |
| Iceland (0.1%) | 66.7% | 33.3% | | |
| Argentina (0.1%) | 50.0% | | | |
| Lithuania (0.1%) | 50.0% | 50.0% | 50.0% | |
| Jersey (0.0%) | | | | |
| Kazakhstan (0.0%) | | | | |

Fig. S6: Percent of companies reducing mortality by at least 1, 10, 100, or 1,000 by country from stated emissions reduction plan. Percentages in parenthesis in each row header denote the percent of all companies analyzed in each country. Percentages in parenthesis in each column header denote percent of all companies exceeding particular threshold.